\documentclass{article}
\usepackage{spconf,amsmath,graphicx}
\usepackage{adjustbox}
\usepackage{color}
\usepackage{amssymb}
\usepackage{url}

\title{AURA-net: robust segmentation of phase-contrast microscopy images with few annotations}

\name{Ethan Cohen$^{\star \dagger}$ \qquad Virginie Uhlmann$^{\star}$}

\address{$^{\star}$ European Bioinformatics Institute, European Molecular Biology Laboratory, Cambridge, UK \\
$^{\dagger}$ \'Ecole Normale Sup\'erieure Paris-Saclay, Paris, France}

\begin{document}
%
\maketitle

\begin{abstract}
We present AURA-net, a convolutional neural network (CNN) for the segmentation
of phase-contrast microscopy images. 
AURA-net uses transfer learning to accelerate training and Attention mechanisms to help the network focus on relevant image features. In this way, it can be trained efficiently with a very limited amount of annotations. 
Our network can thus be used to automate the segmentation of datasets that are generally considered too small for deep learning techniques. AURA-net also uses a loss inspired by active contours that is well-adapted to the specificity of phase-contrast images, further improving performance. We show that AURA-net outperforms state-of-the-art alternatives in several small (less than $100$ images) datasets. 
\end{abstract}


\section{Introduction}
\label{sec:intro}
Label-free imaging denotes a family of non-invasive, non-toxic microscopy techniques that offer an alternative to fluorescence microscopy, particularly for live-cell imaging experiments~\cite{Kasprowicz2017}. Phase-contrast (PC) is one of the oldest label-free imaging modality~\cite{Zernike1942} and remains routinely used in biological experiments. PC uses an optical configuration that translates the phase shifts of light when traversing transparent biological objects into amplitude modulations, resulting in visible differences in image intensity. Objects in PC images exhibit halos and shade-offs, uneven edges appearance, and generally less contrast than in fluorescence microscopy. Imaging biological samples in an undisturbed state, therefore, comes at the cost of images that are more challenging to automatically segment. Segmentation algorithms initially designed for fluorescence microscopy images are indeed observed to be of limited use in PC datasets~\cite{Vicar2019}. 

Active contour (AC) algorithms have been extensively considered for the segmentation of PC images~\cite{Zimmer2002}. AC consist of a curve model that deforms from an initial position in an image by minimizing an energy functional, which can be designed to capture various types of visual features such as regions or edges~\cite{Delgado2014}. The major limitation of AC algorithms is their sensitivity to initial conditions and hyper-parameter settings, restricting their ability to scale and generalize. In recent years, CNN have established themselves as robust alternatives to AC for bioimage segmentation tasks~\cite{Meijering2020}. The ability of CNN to learn relevant features directly from data makes them more robust and allows them to generalize better. Overall, CNN thus tend to be preferred over handcrafted techniques. While many architectures have been proposed in the first years of the deep learning era, U-net~\cite{Falk2019} emerged as an efficient versatile backbone, as exemplified by the many custom versions derived from it~\cite{Xu2019}. Most U-net modifications are however designed to perform best on images featuring objects with strong edges, and few CNN solutions have been dedicated to the segmentation of PC images. Existing solutions either require thousands of ground-truth annotations~\cite{Akram2016}, requesting significant manual labor and restricting their use to large datasets, or rely on fluorescence microscopy data for training~\cite{Ling2020}, thus moving away from the label-free imaging paradigm. Others perform segmentation as an intermediate step towards another analysis goal, such as cell tracking, and therefore only generate rough approximations of cell outlines~\cite{Tsai2019}.

In this work, we present AURA-net, a CNN dedicated to the semantic segmentation of PC microscopy images that only requires a small number of training annotations. Our method integrates elements of AC algorithms with state-of-the-art deep learning strategies: AURA-net combines transfer learning and Attention mechanisms in a U-net architecture trained on a loss inspired by the AC without edges (ACWE) algorithm. The ACWE loss, recently introduced in~\cite{Chen2019}, has been used so far in classical U-net and Dense-net~\cite{Huang2017} architectures. Here, we incorporate it into a more complex CNN and show that its performance outperforms related state-of-the-art CNN methods. 

The paper is organized as follows. In Section~\ref{sec:method}, we introduce the building blocks of our network and describe AURA-net's architecture and loss. We then provide experimental results in Section~\ref{sec:results}, and concluding remarks in Section~\ref{sec:concl}.

\section{Method}\label{sec:method}
\subsection{Building blocks}\label{sec:buildingblocks}
AURA-net's architecture is based on a U-Net integrating transfer learning and Attention mechanisms. In the following, we present in more detail these different building blocks.  

\subsubsection{U-net}
U-net~\cite{Falk2019} is an encoder-decoder CNN composed of a contracting (downsampling) and an expansive (upsampling) path. In the downsampling path, pairs of $3\times 3$ convolution layers are repeatedly applied, followed by a rectified linear unit (ReLU) and a $2\times 2$ max pooling operation with a stride of $2$. The upsampling path then similarly applies blocks of $3\times 3$ convolutions together with upsampling to expand the image back to its original size. U-net uses skip-connections to preserve features at different resolutions and improve localization information. 
Classically, the skip connections apply a concatenation operator between feature maps from the encoder paths and convolution layers in the decoder path, although variants have been proposed. As a result, a large number of features maps are available in the decoder path, allowing information to be transferred efficiently. 

\subsubsection{U-net ResNet}
A popular modification of U-net involves transfer learning~\cite{Yosinski2014}. Transfer learning uses the knowledge a model has acquired in previous training to learn a new related task more efficiently. Using transfer learning in U-Net consists of replacing the downsampling path with a pre-trained deeper network such as ResNet~\cite{He2016}. 
Although available versions of ResNet are typically pre-trained on  ImageNet~\cite{Deng2009}, a database of natural images that significantly differ in appearance from microscopy data, the first layers of the network can detect shapes, colors, and textures at different levels of abstraction and therefore generalize successfully. This type of transfer learning strategy is essential when only a small amount of training data is available. 

\subsubsection{Attention Mechanisms}
Attention~\cite{Vaswani2017} is a popular deep learning concept which aims at identifying discriminating characteristics in internal activation maps, using this knowledge to better represent task-specific data, and ultimately improving the performance of a network. Attention mechanisms help remove less relevant functionalities and focus on more important ones for a given learning task.
Attention gates have been introduced in the U-net architecture and shown to improve its performance in various kinds of biomedical image segmentation tasks~\cite{Oktay2018}.

\subsection{AURA-net}\label{subsec:AURA-Net}
Putting together the concepts described in Section~\ref{sec:buildingblocks}, we propose AURA-net, an {\underline{A}}ttention {\underline{U}}-net {\underline{R}}esNet with an {\underline{A}}C loss. 

\subsubsection{Proposed Architecture}
\begin{figure}
        \centering
    	\includegraphics[width=1.0\linewidth]{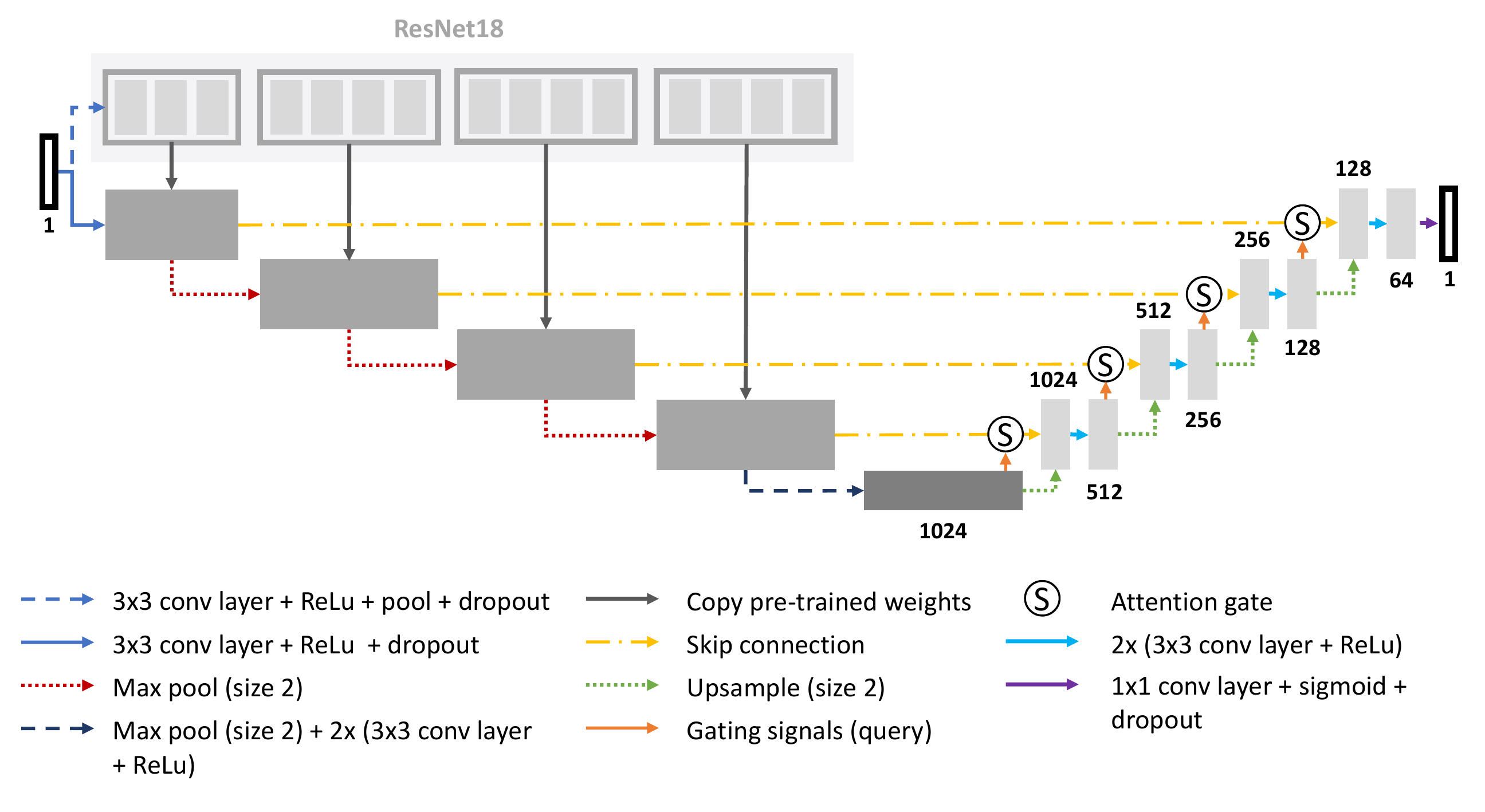}
        \caption{AURA-net's architecture. The network combines a pre-trained ResNet-18 with an Attention-U-net and is trained on an AC loss.}\label{fig:figure}
\end{figure}

The architecture of AURA-net consists of a deeper Attention U-net model, in which we replace the downsampling path with a ResNet-18 model pre-trained on ImageNet. The model is depicted in Figure~\ref{fig:figure}. 
First, the pre-trained ResNet accelerates learning when few annotations are available during training. Second, Attention mechanisms help the network focus on relevant features in poorly contrasted images. Finally, the AC loss captures region information, making up for the faint or absent edges. As a result, our model, combining these strategies in a U-Net architecture, is perfectly suited to the segmentation of PC microscopy datasets when only a small set of annotations is available. 

\subsubsection{Loss Function}
We train AURA-net on a custom loss function composed of three different terms: the well-known pixel-wise binary cross-entropy (BCE)~\cite{Goodfellow2016} and Dice~\cite{Milletari2016} losses, denoted as $\mathcal{L}_\mathrm{BCE}$ and $\mathcal{L}_\mathrm{D}$ respectively, and an AC loss $\mathcal{L}_\mathrm{AC}$ adapted from~\cite{Chen2019}.
The Dice and BCE losses reflect the accuracy of independent pixel-wise predictions but do not enforce region consistency. The addition of the AC term allows incorporating area information, resulting in a loss that combines pixel-wise and region-wise penalties.

The $2$-dimensional AC loss is inspired by the classical ACWE algorithm~\cite{Chan2001} and is defined as
\begin{align}\label{eq:Lac}
       \mathcal{L}_\mathrm{AC}=L+ \lambda A, 
\end{align}
with $\lambda \in \mathbb{R}$. Considering  $I_\mathrm{pred}, I_\mathrm{GT}: \Omega \rightarrow [0,1]$ the predicted and ground truth (reference) image labelling, respectively, $L$ is defined over the image domain $\Omega$ as
\begin{align}\label{eq:L}
    L = \sum_{i,j \in \Omega} \sqrt{|(\nabla_x I_{\mathrm{pred}_{i,j}})^2 + (\nabla_y I_{\mathrm{pred}_{i,j}})^2| + \epsilon},
\end{align}
where $\nabla_n$ denotes the image gradient in direction $n$ and $\epsilon > 0$ is the machine epsilon to avoid numerical errors. This first term captures the length of the
segmentation contour and acts as a regularizer.
The second term, $A$, is expressed as
\begin{align}\label{eq:region}
    A=&\left |\sum_{i,j \in \Omega} I_{\mathrm{pred}_{i,j}} (1 - I_{\mathrm{GT}_{i,j}})^2 \right| \nonumber \\
    &+ \left |\sum_{i,j \in \Omega} (1 - I_{\mathrm{pred}_{i,j}}) I_{\mathrm{GT}_{i,j}}^2 \right|.
\end{align}
It characterizes how closely the predicted objects match ground truth ones in terms of the area they occupy in the image, thus encouraging shape preservation.

We performed line search in the $[0,10]$ interval in order to identify optimal $\lambda$ values. We experimentally observed that the value of $\lambda$ does not strongly influence results as long as it remains larger than $0$, confirming the conclusions of the robustness analysis provided in~\cite{Chen2019}. In practice, we chose to rely on a value of $\lambda=5$ to balance the length and area terms in~$\eqref{eq:Lac}$.

Our final loss function is defined as
\begin{align}\label{eq:loss}
    \mathcal{L} = \gamma \mathcal{L}_\mathrm{AC} + \beta (\alpha \mathcal{L}_\mathrm{BCE}+ (1-\alpha)\mathcal{L}_\mathrm{D}).
\end{align}
The hyper-parameters $\alpha$, $\beta$, and $\gamma$ were chosen as follows. First, $\alpha$ is set to $0.5$ to give equal importance to the Dice and BCE losses. We then relied on line search to retrieve optimal values of $\beta$ and $\gamma$. We identified $\beta=0.75 $ and $\gamma=0.25$ to result in best performance. Note that varying $\alpha \in [0,1]$ while setting $\gamma=0$, $\beta=1$ results in classical pixel-wise segmentation losses.

\section{Experimental Results \& Discussion}\label{sec:results}
\subsection{Datasets}
For our experiments, we rely on three publicly available PC microscopy image datasets from the Boston University Biomedical Image Library (BU-BIL)~\cite{Gurari2015}.
\begin{itemize}
\item Dataset $1$ consists of $35$ raw $1024 \times 811$ images. Each image contains a single rat smooth muscle cell. In our experiments, we randomly selected $25$ images for the training set and $10$ images for the testing set. 
\item Dataset $2$ is composed of $53$ raw images ranging from $278 \times 291$ to $473 \times 659$. Each image contains a single rabbit smooth muscle cell. We randomly split the dataset into $38$ training and $15$ testing images. 
\item Dataset $3$ consists of $48$ raw images of sizes ranging from $356 \times 342$ to $484 \times 434$. Each image contains a single fibroblast.  In our experiments, we randomly picked $36$ images for training and $12$ for testing. 
\end{itemize}
Ground truth binary masks were available for each image for all considered datasets. For each dataset, we resized and cropped images (dataset 1: $512 \times 512$, datasets 2 and 3: $256 \times 256$) for processing. The images were first resized without modifying their aspect ratio such that their smallest dimension matched the target size, and then cropped along their largest dimension.We conducted classical data augmentation, including random flips, rotations, shifts, and scaling, to enrich the training datasets. Besides, we also carried out additional augmentations that are relevant to bioimages, such as shearing, CLAHE~\cite{Pizer1990}, and elastic deformations. We applied the same augmentation techniques for all baselines and for AURA-net.

\subsection{Experimental Strategy}
AURA-net is implemented in PyTorch and is available at {\small\url{github.com/uhlmanngroup/AURA-Net}}.

We quantitatively compared the performance of AURA-net against those of U-net~\cite{Falk2019}, CE-net~\cite{Gu2019}, and Attention U-net~\cite{Oktay2018}. Being the most popular all-around performer for bioimage segmentation, the original U-net and its modified version incorporating Attention mechanisms appeared as natural candidates for comparison. CE-net consists of a modified U-net with pre-trained ResNet blocks in the encoder path and was therefore also a relevant alternative to AURA-net.

All networks were trained on an NVIDIA Tesla K80 with a batch size of $4$, using the Adam optimizer with a learning rate of $3\mathrm{e}{-4}$. We let training run for $100$ epochs for Dataset $1$,  $200$ epochs for Dataset $2$, and $200$ epochs for Dataset $3$. In all cases, we monitored the evolution of the train and validation losses over the epochs to ensure convergence and detect over-fitting. 

\subsection{Results}\label{subsec:results}
\begin{figure}
    \centering
\begin{minipage}[b]{.15\linewidth}
  \centering
      \centerline{\scriptsize Raw}
  \centerline{\includegraphics[width=\linewidth]{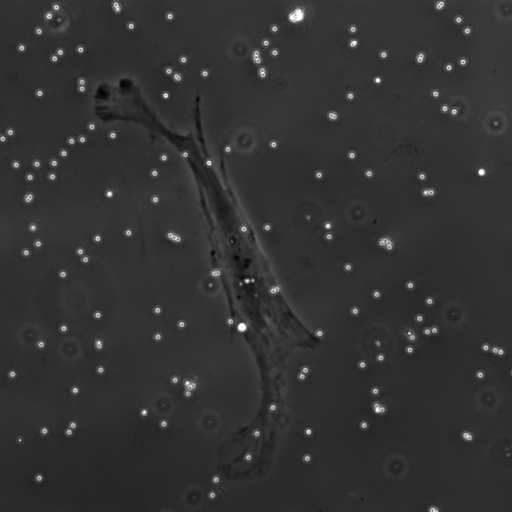}}
\end{minipage}
\begin{minipage}[b]{.15\linewidth}
  \centering
    \centerline{\scriptsize GT}
  \centerline{\includegraphics[width=\linewidth]{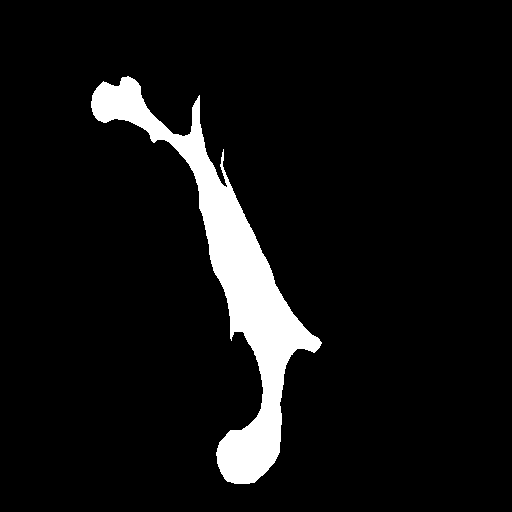}}
\end{minipage}
\begin{minipage}[b]{.15\linewidth}
  \centering
  \centerline{\scriptsize U-net}
  \centerline{\includegraphics[width=\linewidth]{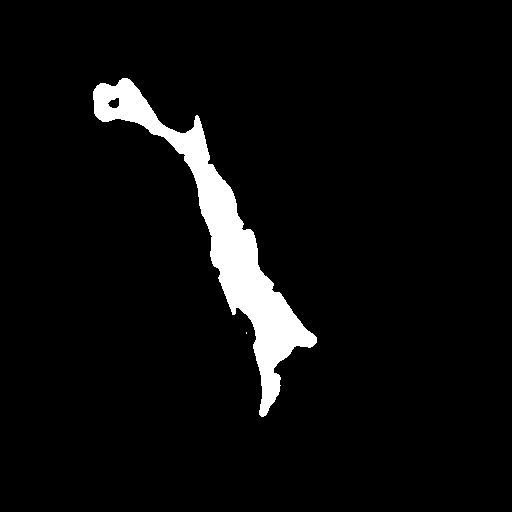}}
\end{minipage}
\begin{minipage}[b]{.15\linewidth}
  \centering
    \centerline{\scriptsize CE-net}
  \centerline{\includegraphics[width=\linewidth]{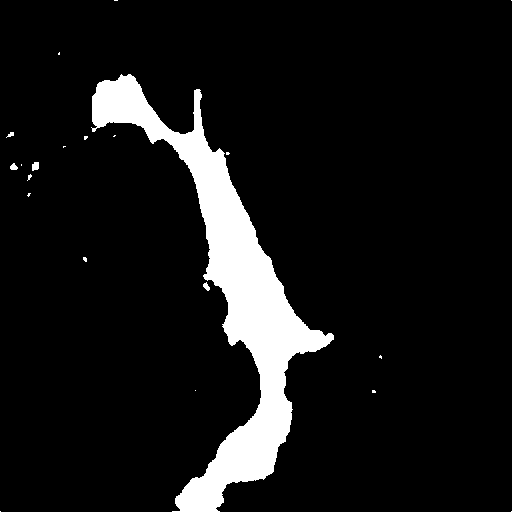}}
\end{minipage}
\begin{minipage}[b]{.15\linewidth}
  \centering
    \centerline{\scriptsize Attention U-net}
  \centerline{\includegraphics[width=\linewidth]{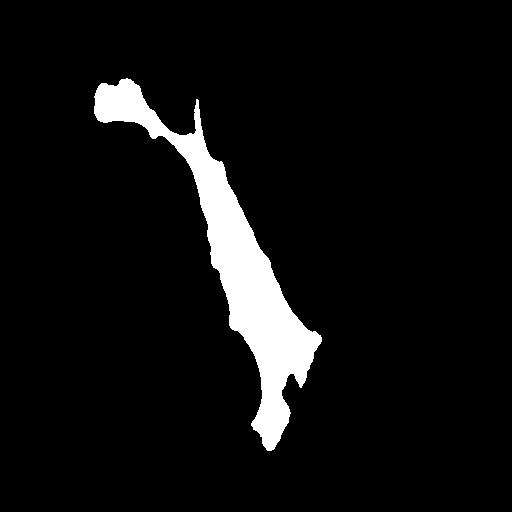}}
\end{minipage}
\begin{minipage}[b]{.15\linewidth}
  \centering
    \centerline{\scriptsize AURA-net}
  \centerline{\includegraphics[width=\linewidth]{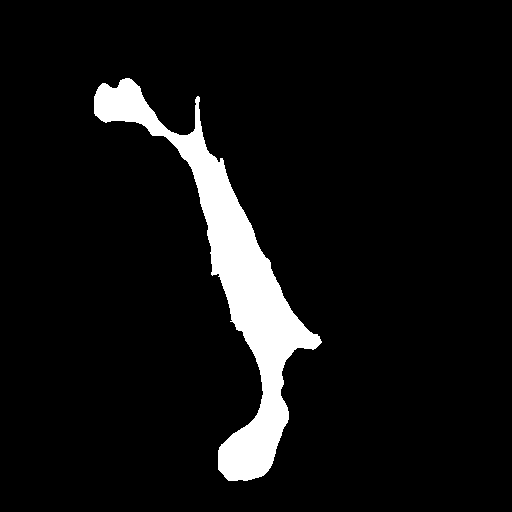}}
\end{minipage}
\begin{minipage}[b]{.15\linewidth}
  \centering
  \centerline{\includegraphics[width=\linewidth]{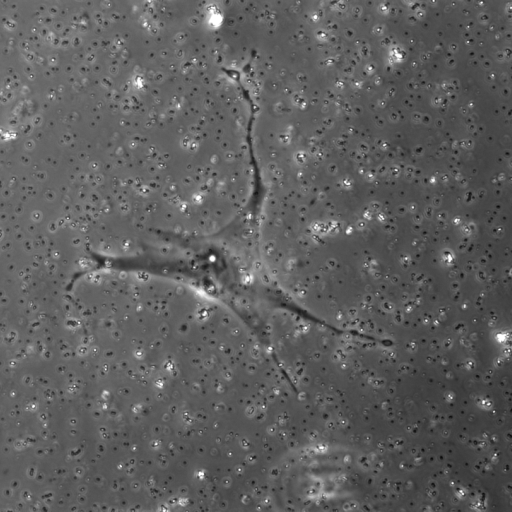}}
\end{minipage}
\begin{minipage}[b]{.15\linewidth}
  \centering
  \centerline{\includegraphics[width=\linewidth]{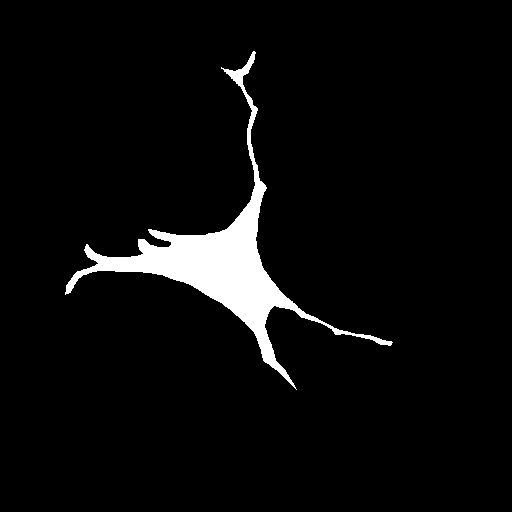}}
\end{minipage}
\begin{minipage}[b]{.15\linewidth}
  \centering
  \centerline{\includegraphics[width=\linewidth]{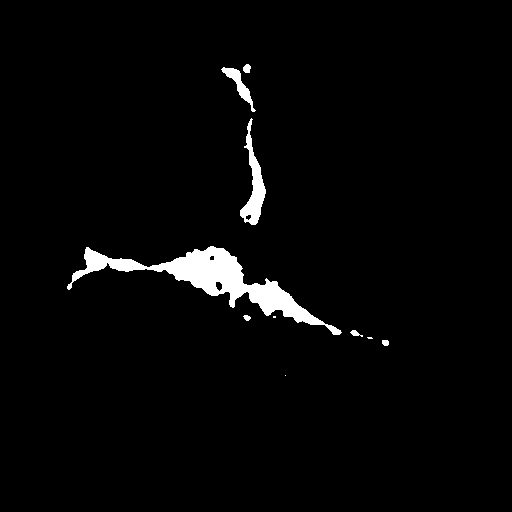}}
\end{minipage}
\begin{minipage}[b]{.15\linewidth}
  \centering
  \centerline{\includegraphics[width=\linewidth]{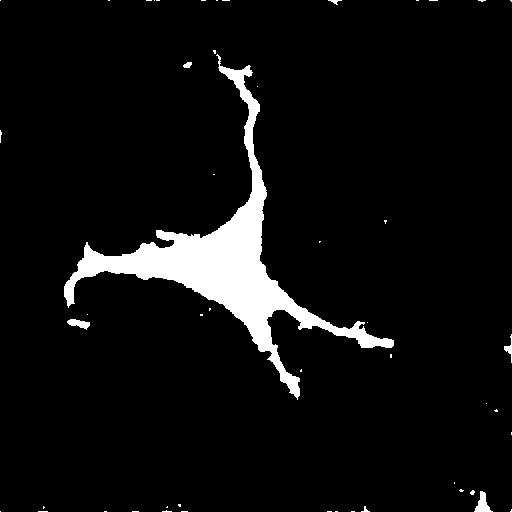}}
\end{minipage}
\begin{minipage}[b]{.15\linewidth}
  \centering
  \centerline{\includegraphics[width=\linewidth]{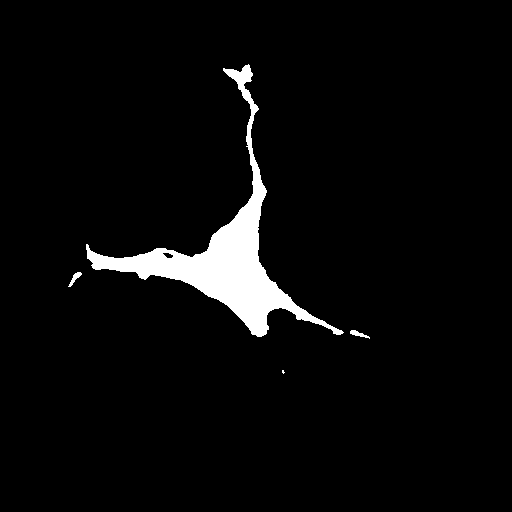}}
\end{minipage}
\begin{minipage}[b]{.15\linewidth}
  \centering
  \centerline{\includegraphics[width=\linewidth]{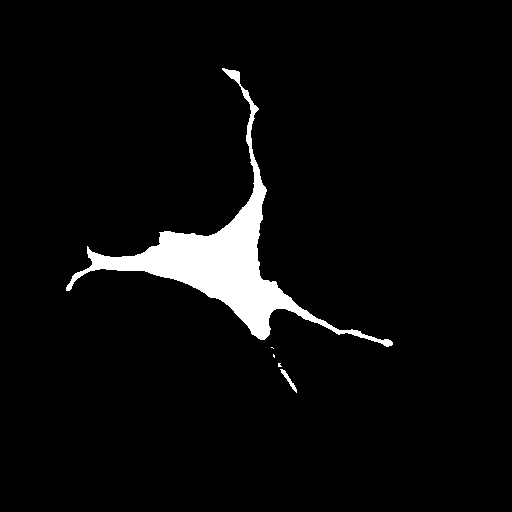}}
\end{minipage}
\begin{minipage}[b]{.15\linewidth}
  \centering
  \centerline{\includegraphics[width=\linewidth]{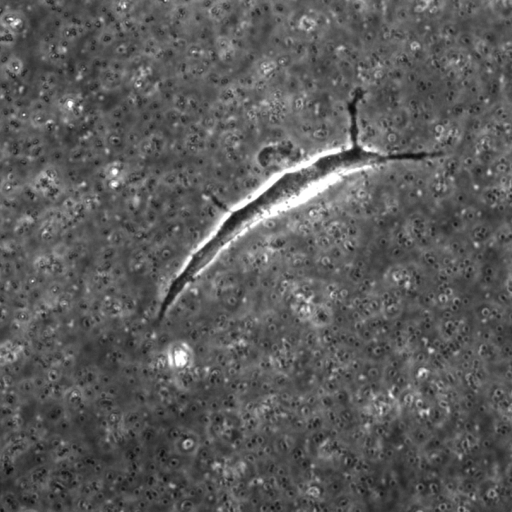}}
\end{minipage}
\begin{minipage}[b]{.15\linewidth}
  \centering
  \centerline{\includegraphics[width=\linewidth]{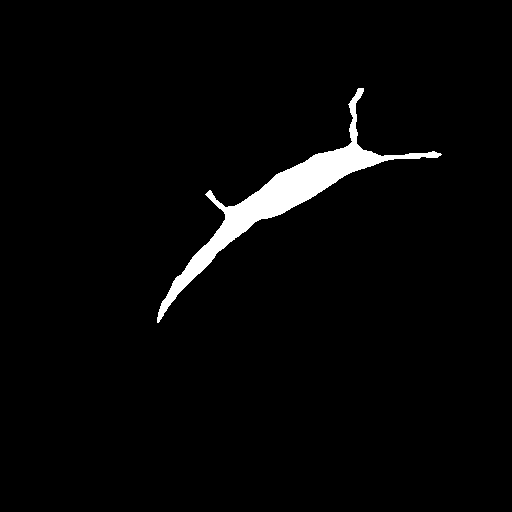}}
\end{minipage}
\begin{minipage}[b]{.15\linewidth}
  \centering
  \centerline{\includegraphics[width=\linewidth]{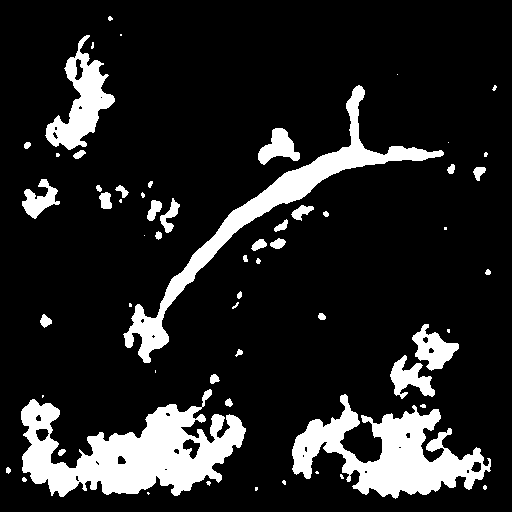}}
\end{minipage}
\begin{minipage}[b]{.15\linewidth}
  \centering
  \centerline{\includegraphics[width=\linewidth]{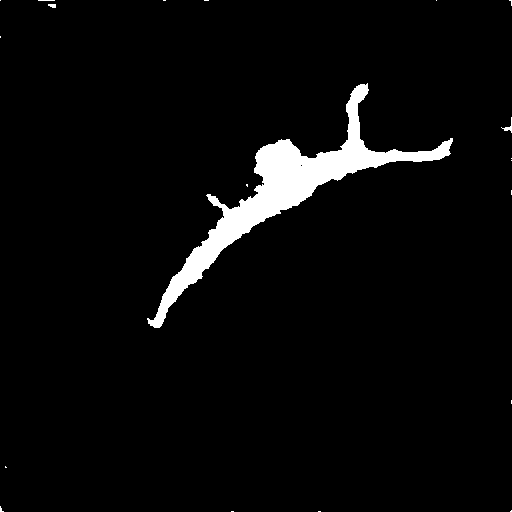}}
\end{minipage}
\begin{minipage}[b]{.15\linewidth}
  \centering
  \centerline{\includegraphics[width=\linewidth]{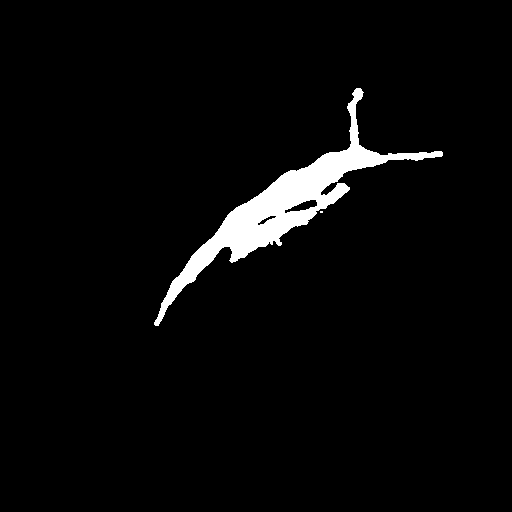}}
\end{minipage}
\begin{minipage}[b]{.15\linewidth}
  \centering
  \centerline{\includegraphics[width=\linewidth]{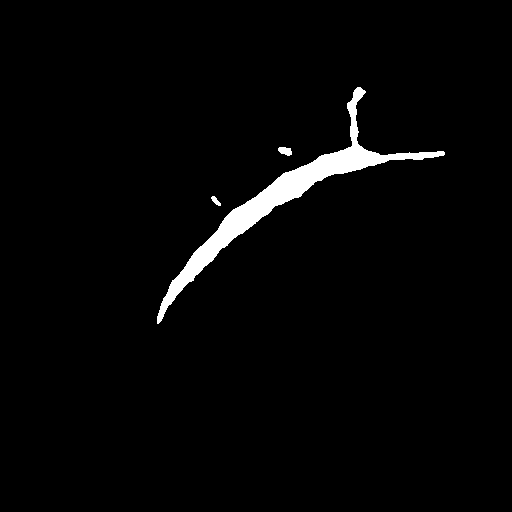}}
\end{minipage}
\caption{Qualitative comparison of AURA-net segmentation results with relevant state-of-the-art alternatives on dataset $1$.}\label{fig:results_1}
\end{figure}

We evaluate the performance of each approach with five classical metrics: intersection over union (IoU), Dice score, precision, recall, and Hausdorff Distance (HD). We provide results on datasets $1$, $2$, and $3$ in Tables~\ref{table:scores}, ~\ref{table:scores_2}, and~\ref{table:scores_3}, respectively. In Figure~\ref{fig:results_1}, we also illustrate representative segmentation results on dataset $1$ for each methods. 

\begin{table}[h!]
\centering
    \begin{adjustbox}{max width=\linewidth}
    \begin{tabular}{l|c|c|c|c|c} 
     & IoU & Dice & Precision & Recall & HD \\
    \hline
    U-net & 36.47\% & 49.92\% & 62.66\% & 63.92\% & 9.71 \\
    CE-net & 59.79\% & 72.87\% & 63.28\% & \textbf{94.23}\% & 8.28 \\
    Attention U-net & 67.44\% & 79.77\% & 82.19\% & 81.57\% & 6.75 \\
    AURA-net &  \textbf{74.64}\% & \textbf{84.61}\% & \textbf{88.17}\% & 83.98\% & \textbf{6.12}
    \end{tabular}
    \end{adjustbox}
    \caption{Comparison of AURA-net with relevant state-of-the-art alternatives on dataset $1$.}
    \label{table:scores}
\end{table}

\begin{table}[h!]
\centering
    \begin{adjustbox}{max width=\linewidth}
    \begin{tabular}{l|c|c|c|c|c} 
     & IoU & Dice & Precision & Recall & HD \\
    \hline
    U-net & 60.74\% & 70.00\% & 72.53\% & 69.78\% & 8.56 \\
    CE-net & 67.67\% & 77.63\% & 70.82\% & 87.46\% & 7.98 \\
    Attention U-net & 75.15\% & 85.14\% & 80.35\% & \textbf{92.01}\% & 5.51 \\
    AURA-net &  \textbf{78.89}\% & \textbf{87.78}\% & \textbf{86.96}\% &  89.87\% & \textbf{ 5.12}
    \end{tabular}
    \end{adjustbox}
    \caption{Comparison of AURA-net with relevant state-of-the-art alternatives on dataset $2$.}
    \label{table:scores_2}
\end{table}

\begin{table}[h!]
\centering
    \begin{adjustbox}{max width=\linewidth}
    \begin{tabular}{l|c|c|c|c|c} 
     & IoU & Dice & Precision & Recall & HD \\
    \hline
    U-net &34.19\% & 45.61\% & 43.95\% & 65.69\% & 5.33 \\
    CE-net & 60.33\% & 74.93\% & 63.05\% & 94.44\% & 3.83 \\
    Attention U-net & 69.62\% & 80.80\% & \textbf{77.95}\% & 85.42\% & 3.55 \\
    AURA-net &  \textbf{69.76}\% & \textbf{81.83}\% & 73.03\% & \textbf{95.31}\% & \textbf{ 3.42}
    \end{tabular}
    \end{adjustbox}
    \caption{Comparison of AURA-net with relevant state-of-the-art alternatives on dataset $3$.}
    \label{table:scores_3}
\end{table}

The original U-net model produces poor segmentation results on datasets $1$ and $3$ while performing better on dataset $2$. CE-net performs better than U-net and results in a strong recall in all three datasets. Attention U-net outperforms both CE-net and U-net on most metrics. The Attention gates incorporated in AURA-net allow outperforming CE-net, while its pre-trained layers help improve over Attention U-net. On top of that, the AC loss provides the network with additional information on object regions. As a result, AURA-net generally outperforms competing approaches. It occasionally scores closely to Attention U-net and CE-net, and even concedes a lead on recall in datasets $1$ and $2$ and on precision in dataset $3$. It however performs best overall, with a consistent clear advantage on the IoU, Dice, and HD metrics.
 
\subsection{Ablation Studies}
Using dataset $1$, we carried out ablation studies to assess the effect of each component of our model. The results (Table~\ref{table:ablation}) illustrate that transfer learning and Attention mechanisms are the most critical elements contributing to the performance of AURA-net. The AC loss further allows the model to improve its performance on all metrics but recall. 

\begin{table}[h!]
\centering
    \begin{adjustbox}{max width=\linewidth}
    \begin{tabular}{l|c|c|c|c|c} 
     & IoU & Dice & Precision & Recall & HD \\
    \hline
    U-net & 36.47\% & 49.92\% & 62.66\% & 63.92\% & 9.71 \\
    U-net+ResNet &  69.96\% &  81.39\% &  83.83\% &  83.71\% & 6.77 \\
    U-net+Attention & 67.44\% & 79.77\% & 82.19\% & 81.57\% & 6.75 \\
    U-net+AC loss &  38.29\%&  53.70\%&  70.50\%&  55.46\%&  11.25\\
    U-net+ResNet+AC loss &   70.31\% & 82.04\% &  83.18\% &  \textbf{84.05}\% & 6.71 \\
    U-net+Attention+AC loss &  68.42\% & 79.16\% &  81.89\% &  80.07\% &  6.65\\
    U-net+ResNet+Attention & 71.27\% &  82.36\% &  84.22\% &  83.90\% & 6.57 \\
    AURA-net &  \textbf{74.64}\% & \textbf{84.61}\% & \textbf{88.17}\% & 83.98\% & \textbf{6.12}
    \end{tabular}
    \end{adjustbox}
    \caption{Ablation studies on dataset $1$.} 
    \label{table:ablation}
\end{table}

\begin{figure}[h!]
        \centering
\begin{minipage}[b]{.25\linewidth}
  \centering
      \centerline{\scriptsize Raw}
  \centerline{\includegraphics[width=\linewidth]{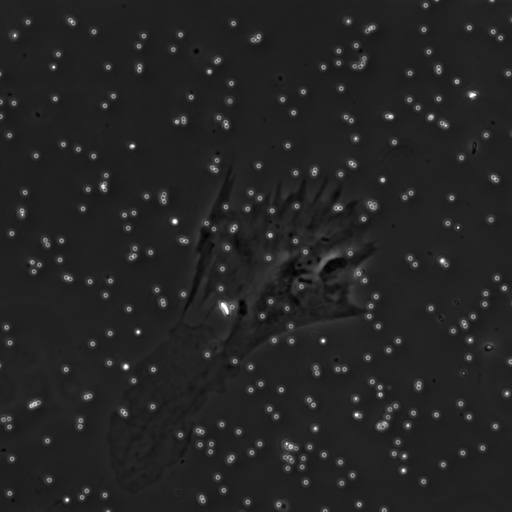}}
\end{minipage}
\begin{minipage}[b]{.25\linewidth}
  \centering
      \centerline{\scriptsize GT}
  \centerline{\includegraphics[width=\linewidth]{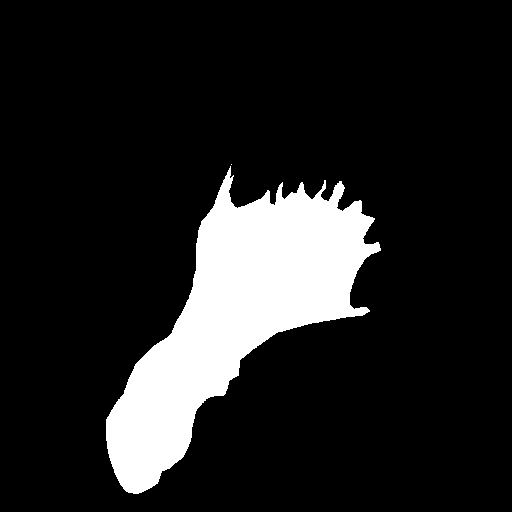}}
\end{minipage} 
\begin{minipage}[b]{.25\linewidth}
  \centering
      \centerline{\scriptsize AURA-net}
  \centerline{\includegraphics[width=\linewidth]{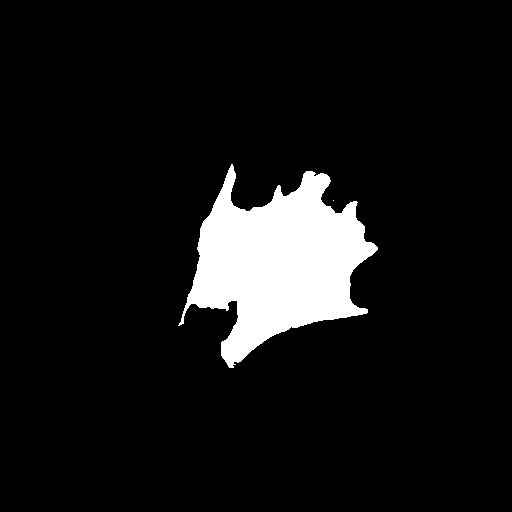}}
\end{minipage}
\\%
\begin{minipage}[b]{.25\linewidth}
  \centering
  \centerline{\includegraphics[width=\linewidth]{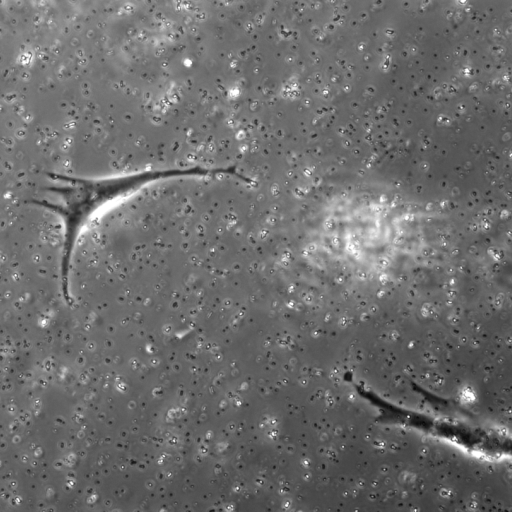}}
\end{minipage}
\begin{minipage}[b]{.25\linewidth}
  \centering
  \centerline{\includegraphics[width=\linewidth]{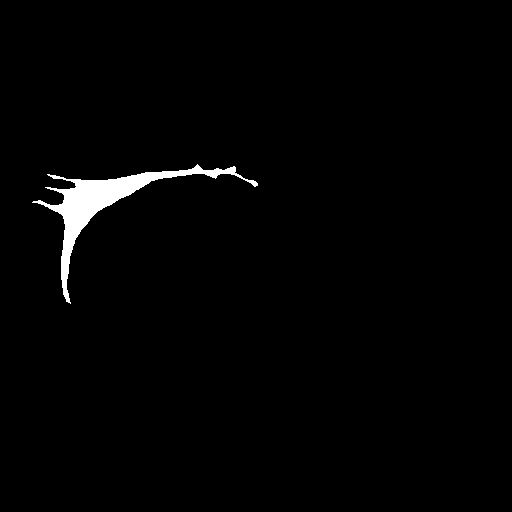}}
\end{minipage}
\begin{minipage}[b]{.25\linewidth}
  \centering
  \centerline{\includegraphics[width=\linewidth]{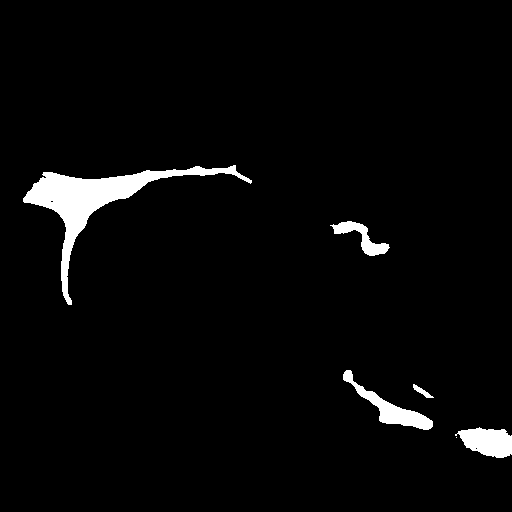}}
\end{minipage}
\caption{Failure cases in dataset $1$.}
        \label{fig:failure_cases}
\end{figure}

\subsection{Failure Cases}
In Figure~\ref{fig:failure_cases}, we illustrate failure cases in dataset $1$. In the first example, AURA-net fails to correctly segment the bottom part of the object. This outcome is unsurprising considering that the raw image exhibits a lower SNR than any of the training data. In the second example, the segmentation mask predicted by AURA-net contains several objects, yielding a poor overlap with the ground truth annotation featuring a single cell. However, the original image reveals the presence of a second, partially cropped non-annotated cell. In this case, part of AURA-net's ``false'' detection are actually correct predictions that have been omitted in ground truth annotations.

\section{Conclusions}\label{sec:concl}
To the best of our knowledge, AURA-net is the first model incorporating state-of-the-art deep learning strategies in a U-net variation targeted to the segmentation of PC microscopy datasets and requiring few training annotations. The qualitative and quantitative results we provide demonstrate that AURA-net is consistently able to accurately segment single cells in PC images relying on only $\sim 30$ training examples. Future work include extending the method to handle images featuring multiple cells in close proximity.
Although we focused on PC, our approach is likely to be useful to segment images acquired with other label-free imaging modalities such as differential interference contrast (DIC) microscopy.

{\small\paragraph*{Acknowledgements} The authors thank Jamie Hackett and Matthieu Boulard for discussions that have inspired this project, as well as Johannes Huger and Julien Fageot for helpful comments on the manuscript. This work was supported by EMBL core fundings and by the EMBL-EBI French Embassy Internship programme. The authors have no relevant financial or non-financial conflict of interest to disclose.
\paragraph*{Compliance with Ethical Standards} This is a computational study for which no ethical approval was required.
}


\bibliographystyle{IEEEbib}
\bibliography{refs}

\end{document}